\documentclass{INTERSPEECH2023}

\usepackage{booktabs,multirow,hyperref,array}
\usepackage{todonotes}
\usepackage{amsmath,graphicx}
\usepackage{amssymb}
\usepackage{lineno}
\usepackage[inline]{enumitem}
\usepackage{xcolor}
\usepackage{verbatim}
\usepackage[hang,flushmargin]{footmisc}

\interspeechcameraready

\title{Evaluating and reducing the distance between \\synthetic and real speech distributions}
\name{Christoph Minixhofer\thanks{\noindent Thanks to Huawei Technologies (UK) Co., Ltd. for funding.\\ We also thank Google's TPU Research Cloud (TRC) for their support.}, Ondřej Klejch, Peter Bell}
\address{
  Centre for Speech Technology Research, University of Edinburgh, United Kingdom
}
\email{\{christoph.minixhofer,o.klejch,peter.bell\}@ed.ac.uk}

\begin{document}

\maketitle
 
\begin{abstract}
 While modern Text-to-Speech (TTS) systems can produce natural-sounding speech, they remain unable to reproduce the full diversity found in natural speech data.  We consider the distribution of all possible real speech samples that could be generated by these speakers alongside the distribution of all synthetic samples that could be generated for the same set of speakers, using a particular TTS system. We set out to quantify the distance between real and synthetic speech via a range of utterance-level statistics related to properties of the speaker, speech prosody and acoustic environment. Differences in the distribution of these statistics are evaluated using the Wasserstein distance. We reduce these distances by providing ground-truth values at generation time, and quantify the improvements to the overall distribution distance, approximated using an automatic speech recognition system. Our best system achieves a 10\% reduction in distribution distance.
\end{abstract}

\section{Introduction}
\label{sec:introduction}

Modern Text-to-Speech (TTS) systems can generate speech close in quality to real speech when evaluated by humans \cite{tan2022naturalspeech,casanova2022yourtts}. Some of these systems even claim to produce synthetic speech indistinguishable from human speech \cite{tan2022naturalspeech}.  However, the ability to produce such synthetic speech hides a limitation of even state-of-the-art TTS systems: the synthetic speech produced is as yet unable to reflect the full diversity of natural speech, even from a fixed set of speakers.  Human speakers can and will produce myriad variations in how they say something, even if the lexical content and intent are the same \cite{Stevens1972}.  In the same conditions, a typical TTS system will produce a single fixed utterance, and with current evaluation methods, a TTS system that can produce ``perfect'' indistinguishable-from-real synthetic speech -- without attempting to addressing any underlying variation -- will be rated highly, even though the synthetic speech may amount to a small subspace of the sample space of real speech.

We believe that increasing this coverage of the real speech distribution by TTS systems could be widely beneficial, for example in improving TTS controllabilty, emotive speech synthesis, or to improve the utility of TTS in as data augmentation for training automatic speech recognition (ASR) system, where capturing as much variation as possible is highly desirable.   An important first step towards this objective is being able to evaluate real speech distribution coverage 
Directly evaluating the distance between the real and synthetic distributions is intractable.  We propose, as a proxy task, to compare the distribution of several speech processing statistics:
\begin{enumerate*}[label=(\arabic*)]
    \item speaker properties, using d-vectors~\cite{wan2018generalized};
    \item prosodic properties, using pitch, energy and speech rate; and
     \item factors related to the acoustic environment, using reverberation and signal-to-noise ratio (SNR). 
\end{enumerate*}
To compare the distributions of the real and synthetic data for the one-dimensional statistics of these proxies, we compute the 2-Wasserstein distance metric \cite{vaserstein1969markov} between distributions.  However, it is still necessary to find an approximate measure of the overall distance between the real and synthetic data distributions, considering that this cannot be readily be evaluated by humans.  In this paper we propose to use an ASR-based method for the evaluation.  It has been shown that an ASR system trained purely on synthetic speech perform very poorly when evaluated on real speech, compared to one trained on the equivalent natural speech \cite{casanova2022single} -- an observation that can be explained by the lower variation in the synthetic training data. We can readily compute a word error rate (WER) ratio between the synthetically-trained system and the naturally-trained system. Since this ratio is 1 for real data and ranges between 3-5 \cite{casanova2022single,rosenberg2019speech} for synthetic data, we believe reducing it is a good proxy for reducing the overall speech distribution distance.

\section{Related Work}

Most work in the TTS domain uses human evaluation in the form of Mean Opinion Scores (MOS) to assess the synthetic speech's quality~\cite{wali2022generative}. However, quantifying naturalness in the generated speech does not evaluate how varied the synthetic speech is compared to the real speech regarding their respective distributions. 

\subsection{Fréchet inception distance}
A metric for this variation in the computer vision domain is the Fréchet inception distance (FID)~\cite{heusel2017gans}, which compares the distribution of generated images with the distribution of the real images.
Efforts have brought this to TTS through the Fréchet DeepSpeech Distance (FDSD) and Kernel DeepSpeech Distance (KDSD) \cite{binkowski2019high}, which measure the distance between distributions of mean and covariances of the last layer representations of DeepSpeech  \cite{hannun2014deep}. However, these measures are not easily interpretable and were introduced as quantitative measures closely correlated with MOS rather than to measure the different effects of divergence in the overall distributions \cite{wali2022generative}.

\subsection{TTS Modeling}
So far, most efforts for bringing the synthetic and real data distributions closer together have introduced improved ways to capture the full data distributions. One group of methods attempts to solve this by making the learning objective more suitable to capture the full diversity of real speech, for example, through a learned latent space \cite{hsu2018hierarchical}, or probabilistic objective \cite{kim2021conditional}. Another group of models extracts features from the target speech and tasks the model to predict them separately, for example, FastSpeech~2 \cite{ren2020fastspeech}, and FastPitch~\cite{lancucki2021fastpitch}. In FastSpeech~2, pitch and energy are predicted for each frame of the target Mel-spectrogram by a "variance adapter". At training time, the ground truth values are quantized and added to the hidden representation input to the decoder. In contrast, values predicted by the variance adapter are used at inference time.

\subsection{Speech Processing measures}
Measures of acoustic environment characteristics most commonly have been used to evaluate the performance of speech enhancement systems. Non-intrusive measures in this field allow for speech waveform evaluation without additional data. Speech-to-Reverberation Modulation Energy Ratio (SRMR) \cite{falk2010non} correlates well with intrusive measures and has become popular in speech enhancement, specifically dereverberation. A non-intrusive estimation of Signal-to-Noise Ratio (SNR) is Waveform Amplitude Distribution Analysis (WADA) \cite{kim2008robust}, although it assumes stationary Gaussian noise. In the speaker domain, there are intra- and inter-speaker variations. In recent multi-speaker TTS systems, d-vector representations derived from speaker-verification systems are frequently used to provide inter-speaker information \cite{stanton2022speaker}. The speaker's current state and their environment can cause intra-speaker variation \cite{Stevens1972}. Differing measures of prosody have been used in speech processing for a long time, with speech rate, pitch and energy being the features used in FastSpeech~2 \cite{ren2020fastspeech}.

\subsection{ASR Evaluation of TTS Speech}
While ASR evaluation of TTS has been applied to compute the WER of the synthesised speech, this previous work does not use this as a measure of the distance between distributions between real and synthetic speech, but rather an alternative to MOS \cite{wali2022generative}. Previous work in this domain has shown that training an ASR system with synthetic data leads to a consistent degradation in Word Error Rate (WER) by a factor of 2 or more \cite{li2018training,rosenberg2019speech,hu2022synt,casanova2022single}. Since for successful ASR model training, much data is needed \cite{lee1988large}, and intra- and inter-speaker variability are crucial \cite{makhoul1995state}, it is worthwhile to find ways to increase this variability. Previous work illustrates this by improving low-resource language ASR by transferring speakers from a high-resource language to a low-resource language using TTS \cite{casanova2022single,chen2021semi,baas2022transfusion}. Explicitly making the distributions of synthetic and real speech data similar regarding speaker variability could allow for even more significant improvements in this area.

\section{Speech Distribution Measures}
\label{sec:distribution}

\begin{figure*}[htb]
\vspace{-.42cm}
\begin{minipage}[b]{1.0\linewidth}
  \centering
  \centerline{\includegraphics[width=0.95\textwidth]{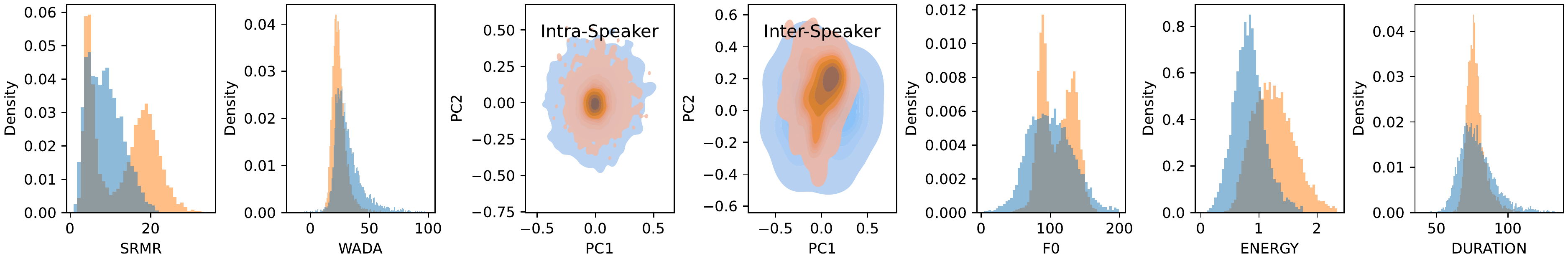}}
  \centerline{(a) measure distributions produced by the \textit{baseline} TTS system}\medskip
\end{minipage}
\begin{minipage}[b]{1.0\linewidth}
  \centering
  \centerline{\includegraphics[width=0.95\textwidth]{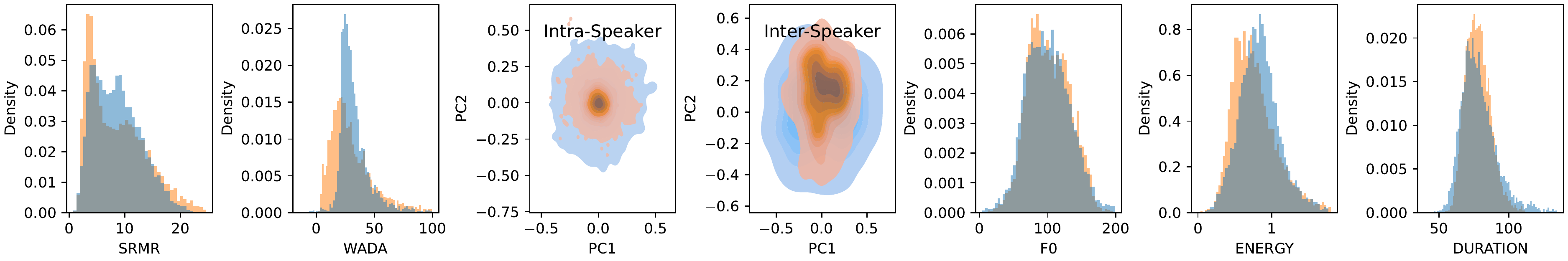}}
  \centerline{(b) measure distributions produced by the TTS system utilizing \textit{measures} and with \textit{acoustic environment augmentation}.}\medskip
\end{minipage}
\caption{Distributions of the real (blue) and synthetic (orange) measures for the \textit{baseline} (top) and improved (bottom) systems.}
\label{fig:res}
\vspace{-.2cm}
\end{figure*}

We first define $\mathcal{X}$ to be the sample space of all possible waveform recordings of speech utterances. We further define $\mathcal{Z}$ as all the possible sets of utterance-level statistics (``metadata'') we can obtain from the waveform -- with and without knowledge of additional contexts such as recording environment or speaker identity -- and $\mathcal{Y}$ as the space possible lexical content. We define $\Omega \subseteq \mathcal{X} \times\mathcal{Y} \times \mathcal{Z}$ as all possible pairs of utterance-level properties, lexical contents and waveforms. A given speech dataset $\mathbb{D}\subset\Omega$ can come with metadata, but in the most common case, it just contains the speaker identity $z$ such that 

\vspace{-.2cm}
$$\mathbb{D}=\{(x^{(i)},y^{(i)},z^{(i)})\}^{|\mathbb{D}|}_{i=1}$$

\subsection{Distance between Real and Synthetic Distributions}
A modern generative TTS model $g_\theta(y,z)$ trained using $\mathbb{D}$ produces a set of synthetic samples with probability distribution $Q(X)$, while the distribution of the real samples is $P(X)$. For the speaker domain, we can compute the fréchet distance (FD) \cite{heusel2017gans} between real and synthetic d-vectors. The FD is defined as the 2-Wasserstein distance between multivariate normal distributions, with an explicit solution \cite{dowson1982frechet}. For the other domains, there are no established measures such as FD. We approximate the divergence between $P(X)$ and $Q(X)$ in different domains by introducing a set of functions $\mathbb{M}$ with any $m: \mathcal{X}\rightarrow \mathbb{R}$, which we refer to as measures. Given their cumulative distribution functions (CDF) $M_Q$ and $M_P$, we can compute the 2-Wasserstein distance \cite{vaserstein1969markov} $W_2$ to quantify the distance between these 1-dimensional distributions for any measure $M$ as follows:
\vspace{-.2cm}
$$W_2(P(M),Q(M)) = \left(\int_\mathcal{X} \left| M_P(x) - M_Q(x) \right|^2 \, \mathrm{d} P(x) \right)^{\frac{1}{2}}$$ 
As shown by \cite{kolouri2018sliced}, $W_2$ can be approximated using the empirical distribution function. If we have $N$ values $m_p$ and $m_q$ from the probability distributions $P(M^{(n)})$ and $Q(M^{(n)})$, respectively and sorted by magnitude, we can approximate $W_2$ as follows:
\vspace{-.2cm}
$$W_2(P(M),Q(M))\approx \frac{1}{N}\left(\sum_{n=1}^{N} \left|m_p^{(n)}-m_q^{(n)}\right|^2\right)^{\frac{1}{2}}$$
This allows us to efficiently approximate the distance between measure probability distributions based on individual samples.

\subsection{Measures}
\label{sec:measures}
Each measure is associated with a domain: the acoustic environment, speaker characteristics or prosody. To quantify the acoustic environment, we use Speech-to-Reverberation Modulation energy Ratio (SRMR) \cite{falk2010non} and Waveform Amplitude Distribution Analysis (WADA) \cite{kim2008robust}, which correlate with the amount of reverberation and noise present in the audio. To quantify speaker characteristics, we rely on the FD between d-vectors obtained from an external speaker verification network \cite{wan2018generalized}, similar to the hidden features used to compute FID in the computer vision domain.
For the intra-speaker case (\textit{FD-Intra}), we subtract the d-vector mean of each speaker, while for the inter-speaker case (\textit{FD-Intra}), we use each speaker's mean d-vector alone. We extract energy and pitch ($F_0$) equally for both the real and synthetic speech. For speech rate ($SR$), we use the average of forced alignment phone durations for real and the durations predicted by $g_\theta$ for the synthetic speech since discrete durations are predicted in the FastSpeech 2 architecture \cite{ren2020fastspeech}. 
The measures' low dimensionality and association with a domain allow us to visualize the synthetic and real distributions, as shown in Figure\ref{fig:res}. We visualize the high-dimensional d-vectors using principal component analysis (PCA). We use $W_2$ as the metric between the real and synthetic samples of any measure -- whenever we do so, we normalize using the ground-truth values first to make measures on different scales comparable.

\subsection{Automatic Speech Recognition}

To make it possible to evaluate how well the individual measures' divergences correspond to the overall speech distribution, we train a speech recognition network on the real and synthetic data independently, resulting in two sets of trained networks. We can then evaluate how well each generalizes to real test data in WER and denote the ratio between the WER achieved using the synthetic data network with the WER of the real data network as the WER-Ratio (\textit{WR}). A lower WER-Ratio correlates with a smaller distance between the real and synthetic data distributions in ASR.

\subsection{Reducing Distribution Distances}
\label{sec:distdist}

We can reduce the distances between the distributions of the previously introduced measures to bring the synthetic and real data distributions closer. In addition to the phoneme sequence, we feed \textit{utterance-level attributes} to the model to allow the TTS model to generate speech with measure distributions more closely matching the ground truth. This expands the metadata $z^{(i)}$ by including speaker d-vectors, as well as mean values for pitch, energy, duration (speaking rate), SRMR and SNR. While this encourages the model not to fall back to generating samples close to the mean for each given measure, we need some way to generate the measures during inference time. Since the space of measures is low-dimensional and varies for each speaker, we use speaker-dependent Gaussian Mixture Models (GMM) for this task. We use a multivariate GMM to generate utterance-specific d-vectors. Factors in the acoustic environment, such as reverberation or background noise, can be hard to generate, but easy to simulate. To test this, we apply both Room Impulse Response (RIR) simulation and additive Gaussian noise as a post-processing step.

\section{Experiments \& Discussion}
\label{sec:experiments}

\begin{table*}[!htb]
\vspace{-.42cm}
\caption{Values of the $W_2$ metric of the different measures, WER Ratio (WR) and WER for our systems.}

\vspace{-.3cm}
  \centering
\begin{tabular}{lccccccccc}
\toprule
\multirow{2}{*}{\textbf{System}} & \multicolumn{2}{c}{\textbf{Speaker}} & \multicolumn{3}{c}{\textbf{Prosody}}         & \multicolumn{2}{c}{\textbf{Environment}} & \multicolumn{2}{c}{\textbf{Overall}} \\
\cmidrule{2-10} 
                                 & \textit{~FD-Intra~} & \textit{~FD-Inter~} & ~~~~$F_0$~~~~         & \textit{Energy}          & ~~~$SR$~~~           & \textit{~~SRMR~~}                 & \textit{WADA SNR}                 & \textit{~~~~WER~~~~}       & \textit{~~~WR~~~}     \\
\midrule
Baseline                                   & 0.36              & 0.57              & 0.57          & 1.97          & 0.21          & 1.58                & 0.57                & 48.6$\pm0.43$          & 3.66$\pm0.03$          \\
+ Environment                                & 0.33              & 0.93              & 0.13          & 1.35          & 0.17          & 1.33                & 0.56                & 49.2$\pm0.51$          & 3.70$\pm0.04$          \\
+ Attributes                                 & 0.30              & \textbf{0.50}     & 0.06          & 0.15          & 0.15          & 1.23                & 0.58                & 47.2$\pm0.39$          & 3.55$\pm0.03$          \\
+ Augmentation                               & 0.44              & 0.66              & \textbf{0.02} & \textbf{0.08} & 0.15          & \textbf{0.05}       & \textbf{0.23}       & 44.0$\pm0.33$          & 3.31$\pm0.02$          \\ 
\midrule
Oracle                                     & \textbf{0.17}     & 0.65              & 0.09          & 0.81          & \textbf{0.09} & 0.16                & 0.27                & \textbf{43.0}$\pm0.29$ & \textbf{3.24}$\pm0.02$ \\
\midrule
Real Data & \textit{-}        & \textit{-}        & \textit{-}    & \textit{-}    & \textit{-}    & \textit{-}          & \textit{-}          & \textit{13.3}             & \textit{-} \\ 
\bottomrule
\end{tabular}

\label{tab:srwf}
\vspace{-.2cm}
\end{table*}

\begin{figure}[h]
\vspace{-.2cm}
  \centering
  \centerline{\includegraphics[width=0.6\linewidth]{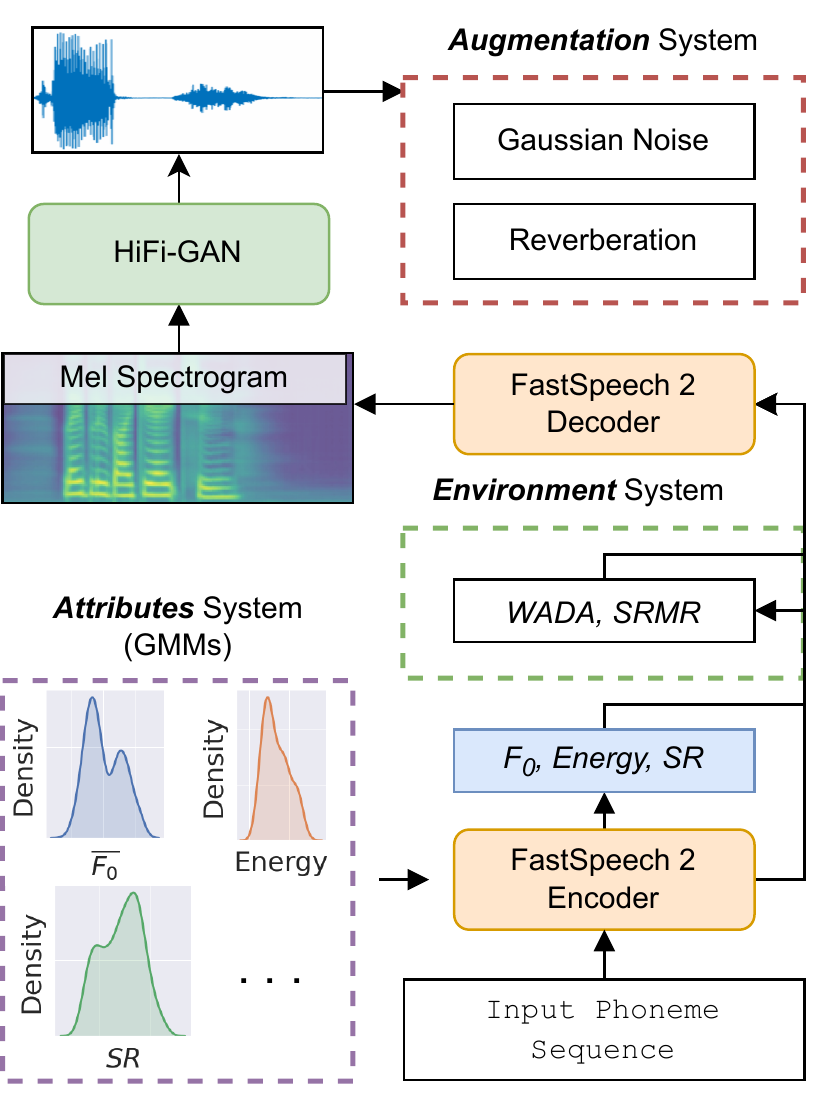}}
  \caption{\textbf{\textit{Environment}}, \textbf{\textit{Attributes}} and \textbf{\textit{Augmentation}} systems.}
  \label{fig:system}
  \vspace{-.6cm}
\end{figure}

For our experiments, we compute the $W_2$ metric of the different measures introduced in Section~\ref{sec:measures} for different TTS systems. These systems aim to reduce the distance between real and synthetic speech distributions. To explore if the distance between the distributions is reduced, we compute the WER and WR (WER Ratio) achieved using a 6-layer hybrid HMM-TDNN system \cite{peddinti2015time} with a hidden unit size of 512, trained with the LF-MMI \cite{povey2016purely} objective for 4 epochs (3-4 hours on 4 GTX 1080 Ti). We keep our changes to the ASR system minimal to quantify the synthetic-real distribution distance with little interference from the ASR system. Our TTS systems were trained for 40 epochs, which took $\approx$ 24 hours on 4 GTX 1080 Ti GPUs.

\subsection{Dataset}
For all our experiments, we use the clean split of LibriTTS. We only use speakers with at least 100 utterances, which leaves us with 684 speakers. Half of each speaker's utterances are reserved for inference, and the other half is the training data for the TTS system -- this way, we can generate a synthetic corpus paired with real speech samples. We generate 10 hours of audio for each of our systems, with equal transcripts and speakers.

\subsection{TTS Systems}
We use FastSpeech~2  with the efficiency improvements suggested in~\cite{luo2021lightspeech}.
While the original system only allows single-speaker use, we make it suitable for multi-speaker training using d-vectors extracted from utterances and averaged to arrive at one d-vector per speaker. We extract d-vectors from a speaker verification model
~trained on the VoxCeleb1 dataset \cite{Nagrani17}. We then add the d-vectors to the hidden sequence before the encoder and decoder. To help model many speakers, we increase the number of decoder layers from 4 to 6. Rather than the Continous Wavelet Transform (CWT) used for Pitch in FastSpeech~2, we directly predict frame-level pitch and energy contours using Soft-DTW loss \cite{cuturi2017soft}. As in \cite{chien2021investigating} we use HiFi-GAN to produce raw waveforms at inference time. Additionally, we use VoiceFixer \cite{liu2021voicefixer} to remove unwanted artefacts from the synthetic speech. This 10.4M parameter model serves as our \textit{baseline} system. We incrementally add techniques to this system to bring the real and synthetic distributions closer. 
For the \textit{\textbf{Environment}} system, we add frame-level WADA and SRMR features to the ``variance adaptor'', tasking the model to predict values related to the acoustic environment in addition to prosody. At inference time, we sample attributes from speaker-specific GMMs, as described in Section~\ref{sec:distdist}. In line with the ``variances'' used in FastSpeech~2~\cite{ren2020fastspeech}, we quantize each measure into one of 256 possible values and convert them into measure embedding vector $m^{(i)}$, which we add to the expanded hidden sequence after the last layer of the encoder.
For the \textit{\textbf{Attributes}} system, the model is additionally supplied with the ground-truth value of each measure introduced in Section \ref{sec:measures}, which are sampled from GMMs. In preliminary experiments, we find two components for each GMM with a variance floor of $10^{-3}$ to be a robust choice.
For the \textit{\textbf{Augmentation}} system, we further reduce the distance in terms of acoustic environment. We add Gaussian noise with a target SNR ranging from 5 dB to 40 dB. We also add an RIR with a probability of $0.8$ with target $RT_{60}$ (reverberation time for a 60 $dB$ signal) \cite{Eargle2013-ny} ranging from 0.15 to 0.8. Both values are only randomized once per speaker. We use the audiomentations library\footnote{\href{https://github.com/iver56/audiomentations}{https://github.com/iver56/audiomentations}}.
We compare our systems to an \textit{\textbf{Oracle}} system where the ground-truth values for all attributes are known, and augmentation is applied. This allows us to quantify how effective the GMMs are at producing realistic attributes. A big difference in the metrics between the oracle system and our systems would suggest a failure of the GMMs to produce realistic attributes, rather than a failure of the TTS system itself.
The aforementioned systems are illustrated in Figure~\ref{fig:system}.

As shown in Table~\ref{tab:srwf}, the baseline system has a WR of 3.66, meaning the synthetic speech is more than 3 times less effective than real speech for ASR training. Once all the above improvements are applied, the WR drops to 3.31, a 10\% reduction. In the following sections, we discuss the distribution distances and overall distance with respect to different domains.

\vspace{-.2cm}
\subsection{Speakers}
\label{sec:speakers}
As expected the \textit{\textbf{Attributes}} system to reduce the intra-speaker distribution distance (\textit{FD-Intra}), since the sampling of utterance-level d-vectors should increase intra-speaker variety. There is a large reduction in \textit{FD-Intra} (16.6\%). The \textit{\textbf{Attributes}} system also improves \textit{FD-Inter} distance (12.2\%). While this is unexpected, we hypothesise it is due to a better match between d-vectors and individual utterances during training. Both effects are diminished when used with \textit{\textbf{Augmentation}} but still lead to a reduced \textit{WR}. The speaker distribution distances are the only domain in which the \textit{\textbf{Oracle}} system is closer to the real data distribution across the board, especially for the inter-speaker case. We hypothesise this is due to the multivariate GMM for speaker generation being too limited.

\vspace{-.2cm}
\subsection{Prosody}
\label{sec:prosody}
As shown in Figure \ref{fig:res}a, the synthetic distributions for $F_0$, \textit{Energy} and $SR$ distributions diverge from their real counterparts differently. $F_0$ shows two peaks as opposed to the normally distributed real distribution, while the energy distribution \textit{Energy} is shifted, and the duration distribution $SR$ shows a smaller standard deviation than its real counterpart. The \textit{\textbf{Attributes}} TTS system mitigates this while also having a positive effect on \textit{WR}. Interestingly, the \textit{\textbf{Oracle}} system shows a high distribution distance for the energy measure while maintaining a low \textit{WR}. This suggests that energy does not strongly contribute to speech distribution distance.

\vspace{-.2cm}
\subsection{Acoustic Environment}
\label{sec:rec}
We find the acoustic environment important for the overall distribution distance approximated by \textit{WR}. The \textit{\textbf{Augmentation}} model reduces \textit{WR} by 6.7\%, which is the largest reduction of any system. However, it slightly increases both intra- and inter-speaker distribution distance. Adding environmental measures to the prediction task in FastSpeech 2 is less effective, slightly reducing SRMR but increasing intra-speaker distribution distance. Overall, we note a minor increase in WR (1\%); however, with \textit{\textbf{Attributes}}, \textit{SRMR} is reduced significantly (22.1\%). We do not observe a significant improvement to \textit{WADA SNR} - we hypothesise the model might be incapable of producing the Gaussian noise quantified by the WADA SNR algorithm.

\vspace{-.2cm}
\section{Conclusion and Future Work}
\label{sec:conclusion}

As TTS systems creating synthetic speech indistinguishable from real speech are within reach, we cast our eye towards variety in synthetic speech. This variety can be expressed as how closely the synthetic data distribution matches the real data distribution. To quantify this, we have introduced a way to measure the distance between real and synthetic speech distributions using a range of measures of the acoustic environment, speakers and prosody domains. We introduced a metric approximating the distance between real and synthetic speech using ASR, the WER ratio (\textit{WR}). We have shown that for a multi-speaker version of the non-autoregressive TTS model FastSpeech~2, prosody and speaker distribution distance can be effectively reduced by conditioning on utterance-level measures during training which are sampled from GMMs at inference time. While said GMMs are very effective at modelling prosody measures, they are less effective than an oracle system for speaker distribution distances. Additionally, we have shown that the mismatches between acoustic environments can be rectified using a simple data augmentation strategy.




\bibliographystyle{IEEEtran}
\bibliography{mybib}

\end{document}